\documentclass[%
 reprint,
twocolomn,
 amsmath,amssymb,
 aps,
pra,
]{revtex4-2}

\usepackage{graphicx}
\usepackage{dcolumn}
\usepackage{bm}
\usepackage{bm}
\usepackage{color}

\begin{document}


\title{Optimized many-hypercube codes toward lower logical error rates and earlier realization}

\author{Hayato Goto$^{1,2,}$}
 \email{hayato.goto@riken.jp}

\affiliation{%
$^1$RIKEN Center for Quantum Computing (RQC), Wako, Saitama 351-0198, Japan \\
$^2$Corporate Laboratory, Toshiba Corporation, Kawasaki, Kanagawa 212-8582, Japan}%

\date{\today}

\begin{abstract}
Many-hypercube codes, 
concatenated ${[[n,n-2,2]]}$ quantum error-detecting codes ($n$ is even), 
have recently been proposed as high-rate quantum codes suitable for fault-tolerant quantum computing. 
While the original many-hypercube codes with ${n=6}$ can achieve remarkably 
high encoding rates 
(about 30\% and 20\% at concatenation levels 3 and 4, 
respectively), 
they have large code block sizes at high levels (216 and 1296 physical qubits per block at levels 3 and 4, respectively), 
making not only experimental realization difficult but also logical error rates
per code block high. 
Toward earlier experimental realization and lower logical error rates, 
here we comprehensively investigate smaller many-hypercube codes with $[[6,4,2]]$ and/or $[[4,2,2]]$ codes, where, e.g., $D_{6,4,4}$ denotes the many-hypercube code using $[[6,4,2]]$ at level 1 and $[[4,2,2]]$ at levels 2 and 3. As a result, we found a counterintuitive fact that $D_{6,4,4}$ ($D_{6,6,4,4}$) can achieve lower logical error rates per code block than $D_{4,4,4}$ ($D_{4,4,4,4}$), despite its higher encoding rate and larger code block size. 
Focusing on level 3,
we also developed efficient fault-tolerant encoders realizing about 60\% overhead reduction while maintaining or even improving the performance, 
compared to the original design.
Using them, we numerically confirmed that  
$D_{6,4,4}$ also achieves the best performance for logical controlled-NOT gates 
in a circuit-level noise model. 
These results are important for targeting a high-rate code toward early experimental realization of efficient fault-tolerant quantum computing.
\end{abstract}

\maketitle


\section{Introduction}

Building a reliable quantum computer is challenging, 
because of noises, e.g., due to decoherence and 
imperfection of quantum-system control.
To overcome this difficulty, 
quantum error-correcting codes~\cite{Shor1995a,Steane1996a,Calderbank1996a,Steane1996b,Steane1996c,Laflamme1996a,Gottesman1996a} and 
fault-tolerant quantum computation (FTQC) with them~\cite{Shor1996a,Gottesman1998a,Nielsen} 
have been proposed. 
Since the threshold theorem 
that arbitrarily long quantum computations 
can be performed reliably if physical error rates are lower than a threshold was proved 
by using concatenated quantum codes~\cite{Aharonov1997a},
FTQC had been based on concatenated quantum codes~\cite{Knill1996a,Zalka1996a,Knill1998a,Steane2003a,Knill2005a,Aliferis2006a,Goto2009a,Goto2010a,Fujii2010a,Goto2013a,Goto2014a,Goto2014b}.
This approach culminated by Knill's ${C_4/C_6}$ scheme showing the threshold over 1\%, 
which is based on the concatenation of simple quantum error-detecting codes~\cite{Knill2005a}.
At that time, however, 
it was extremely difficult to experimentally realize these proposals, 
in particular, nonlocal gates 
such as transversal controlled-NOT (CNOT) gates~\cite{Nielsen}.

Topological codes~\cite{Kitaev1996a}, 
in particular, the surface code~\cite{Bravyi1998a,Dennis2002a},  
have become more popular 
since it was shown that 
the surface code can achieve 
comparable thresholds only using local operations~\cite{Raussendorf2007a,Raussendorf2007b,Fowler2009a,Wang2011a,Fowler2012a}.
Moreover, a compact version called the rotated surface code and a simple two-qubit gate approach called lattice surgery 
were also proposed~\cite{Horsman2022a}.
The surface-code approach to FTQC is particularly suitable for 
superconducting-circuit implementations, 
leading to recent experimental demonstrations 
of the surface code with superconducting circuits~\cite{Krinner2022a,Zhao2022a,Google2023a,Google2024a}.

The (rotated) surface code encodes 
a single logical qubit 
into $d^2$ physical qubits, 
where $d$ is the code distance.
Thus, the encoding rate $R=1/d^2$ 
becomes low for large code block sizes, 
resulting in large resource overheads.
To address this overhead issue, 
quantum low-density parity-check (qLDPC) codes~\cite{Breuckmann2021a,Gottesman2014a,MacKay2004a,Kovalev2013a,Leverrier2015a,Fawzi2018a,Breuckmann2021a,Panteleev2022a,Dinur2023a,Leverrier2022a,Lin2022a}, which can achieve 
a constant encoding rate even for large code distances, 
have recently attracted much attention~\cite{Gottesman2022a,Tremblay2022a,Viszlai2023a,Bravyi2024a,Xu2024a,Poole2024a}. 
This expectation is enhanced by recent experimental advances, in particular, in ion-trap~\cite{Pino2021a,Ryan2021a,Postler2022a,Moses2023a,Wang2024a,Postler2024a,Self2024a,Yamamoto2024a,Silva2024a,Ryan2024a,Dasu2025a,Daguerre2025a,Hughes2025a,Ransford2025a} and 
neutral-atom~\cite{Graham2022a,Bluvstein2022a,Evered2023a,Bluvstein2024a,Muniz2025a,Bluvstein2025a,Rodriguez2025a,Chiu2025a,Manetsch2025a,Infleqtion2025a} platforms, 
because long-range connectivity beyond nearest-neighbor one necessary for high-rate codes such as qLDPC codes~\cite{Bravyi2010a,Baspin2022a} is experimentally feasible by moving qubits in these platforms.
However, parallel execution of addressable logical gates with low overheads is challenging for qLDPC codes, 
the study on which is ongoing~\cite{Krishna2021a,Cohen2022a,Quintavalle2023a,Williamson2024a,Swaroop2024a,Cross2024a,Patra2024a,Malcolm2025a,Zhang2025a,Xu2025a,Ide2025a,Cowtan2025a,He2025a,Yoder2025a,Baspin2025a,Xu2025b,Zheng2025a,Cowtan2025b,Zhang2025b,Pecorari2025a,Webster2025a,Tan2025a,Tamiya2025a}.

Alternative high-rate codes, namely, non-LDPC high-rate quantum codes 
have also been proposed toward low-overhead FTQC~\cite{Yamasaki2024a,Goto2024a,Yoshida2025a,Litinski2025a,Liu2025a,Kanomata2025a,Nakai2025a,Berthusen2025a}.
Among them, concatenated high-rate quantum codes are particularly promising 
in the sense that fault-tolerant logical-state preparation can be achieved by level-by-level encoding~\cite{Knill2005a,Goto2014a,Dasu2025a,Yamasaki2024a,Goto2024a,Yoshida2025a}.
In this work, we focus on concatenated ${[[n,n-2,2]]}$ quantum error-detecting codes ($n$ is even), 
which are called many-hypercube (MHC) codes because their structures can be described by using 
many $L$-dimensional hypercubes ($L$ is the concatenation level) 
corresponding to logical qubits (logical Pauli operators)~\cite{Goto2024a,Liu2025a,Nakai2025a}.
Note that the error-detecting codes, which are also called iceberg codes~\cite{Self2024a}, 
are the simplest, highest-rate Calderbank-Shor-Steane (CSS) codes~\cite{Calderbank1996a,Steane1996b,Steane1996c}.
In the original work~\cite{Goto2024a}, 
$[[6,4,2]]$ was adopted as the base code 
because of its good balance between the encoding rate and the code block size. 
Recently, smaller MHC codes with $[[4,2,2]]$ have been studied~\cite{Liu2025a,Nakai2025a}.
Remarkably, the smallest level-2 MHC code $[[4^2,2^2,2^2]]$ has been realized experimentally using neutral atoms~\cite{Infleqtion2025a}.
Smaller block sizes will lead to not only earlier experimental realization, 
but also lower logical error rates per code block (block error rates), 
though the encoding rates become lower.
For this reason, in this work we comprehensively investigate the MHC codes with $[[6,4,2]]$ and/or $[[4,2,2]]$, 
where the MHC code using ${[[n_L,n_L-2,2]]}$ at level $L$ is 
denoted by $D_{n_1,n_2,\ldots}$~\cite{Liu2025a}.
As a result of code capacity experiments, we found a counterintuitive fact that $D_{6,4,4}$ ($D_{6,6,4,4}$) can achieve lower block error rates than $D_{4,4,4}$ ($D_{4,4,4,4}$), despite its higher encoding rate and larger code block size. 
In addition, it turned out that $D_{4,4,6,6}$ is the worst at level 4.
This is a notable result, because in a previous study~\cite{Liu2025a}, 
it was assumed that we should use smaller codes at lower levels, 
and consequently $D_{4,4,6,6}$ was chosen in the previous study.
We also develop efficient fault-tolerant encoders for level-3 MHC codes, 
which can reduce overheads by about 60\% compared to the original design~\cite{Goto2024a}.
Using them, we confirmed that $D_{6,4,4}$ can also achieve the highest performance at level 3 
in a circuit-level noise model.

\section{Definitions of MHC codes}
\label{sec2}

We first present the detailed definitions of the MHC codes from level 1 
to level 4.
We will use figures to show the code structures, 
but note that they are conceptual and we do not have to realize 
such qubit configurations in real space.

\subsection{$D_4$ and $D_6$}

In this work, we use 
${[[ 4,2,2]]}~D_4$ and/or ${[[ 6,4,2]]}~D_6$ 
as the base codes for the MHC codes.
Both have only two stabilizers, all $Z$ and all $X$, 
and their logical Pauli operators have weight of 2, 
as shown explicitly below, 
leading to code distance of 2.

\begin{figure*}[t]
    \includegraphics[width=1.3\columnwidth]{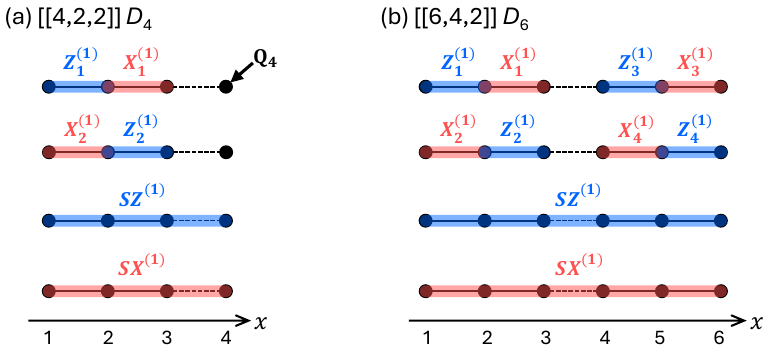}
    \caption{Definitions of (a) $D_4$ and (b) $D_6$. 
    The circles represent physical qubits. 
    The top two show logical Pauli operators and the bottom two show stabilizers.
    $Z$ and $X$ operators are shown in blue and red, respectively.
    Solid and dashed line segments connecting qubits are guides to the eye.}
    \label{fig1}
\end{figure*}

The logical Pauli operators and stabilizers for $D_4$ are defined as follows:
\begin{align}
Z_{1}^{(1)}&=Z_1 Z_2,& X_{1}^{(1)}&=X_2 X_3, \nonumber\\
Z_{2}^{(1)}&=Z_2 Z_3,& X_{2}^{(1)}&=X_1 X_2, \nonumber\\
SZ^{(1)}&=Z_1 Z_2 Z_3 Z_4,&
SX^{(1)}&=X_1 X_2 X_3 X_4, \nonumber
\end{align}
where $Z_i$ and $X_i$ are the Pauli operators for the $i$th physical qubit, 
which is denoted by $\mathrm{Q}_i$, 
and the superscript suggests that 
they are the operators used at level 1 of MHC codes.
By aligning the four physical qubits along the virtual $x$ axis, 
where $\mathrm{Q}_i$ is placed at $x=i$, 
the above operators can be visualized as shown in Fig.~\ref{fig1}(a).

The logical Pauli operators and stabilizers for $D_6$ are defined as follows:
\begin{align}
Z_{1}^{(1)}&=Z_1 Z_2,& X_{1}^{(1)}&=X_2 X_3, \nonumber\\
Z_{2}^{(1)}&=Z_2 Z_3,& X_{2}^{(1)}&=X_1 X_2, \nonumber\\
Z_{3}^{(1)}&=Z_4 Z_5,& X_{3}^{(1)}&=X_5 X_6, \nonumber\\
Z_{4}^{(1)}&=Z_5 Z_6,& X_{4}^{(1)}&=X_4 X_5, \nonumber\\
SZ^{(1)}&=Z_1 Z_2 Z_3 Z_4 Z_5 Z_6,&
SX^{(1)}&=X_1 X_2 X_3 X_4 X_5 X_6. \nonumber
\end{align}
Similarly to the case of $D_4$, 
these can be visualized as shown in Fig.~\ref{fig1}(b).
There are different definitions of $[[n,n-2,2]]$ codes~\cite{Self2024a}, 
but we adopt the above definitions because of the simple visualization~\cite{Goto2024a}.

\subsection{Level-2 MHC codes}

\begin{figure*}[t]
    \includegraphics[width=1.3\columnwidth]{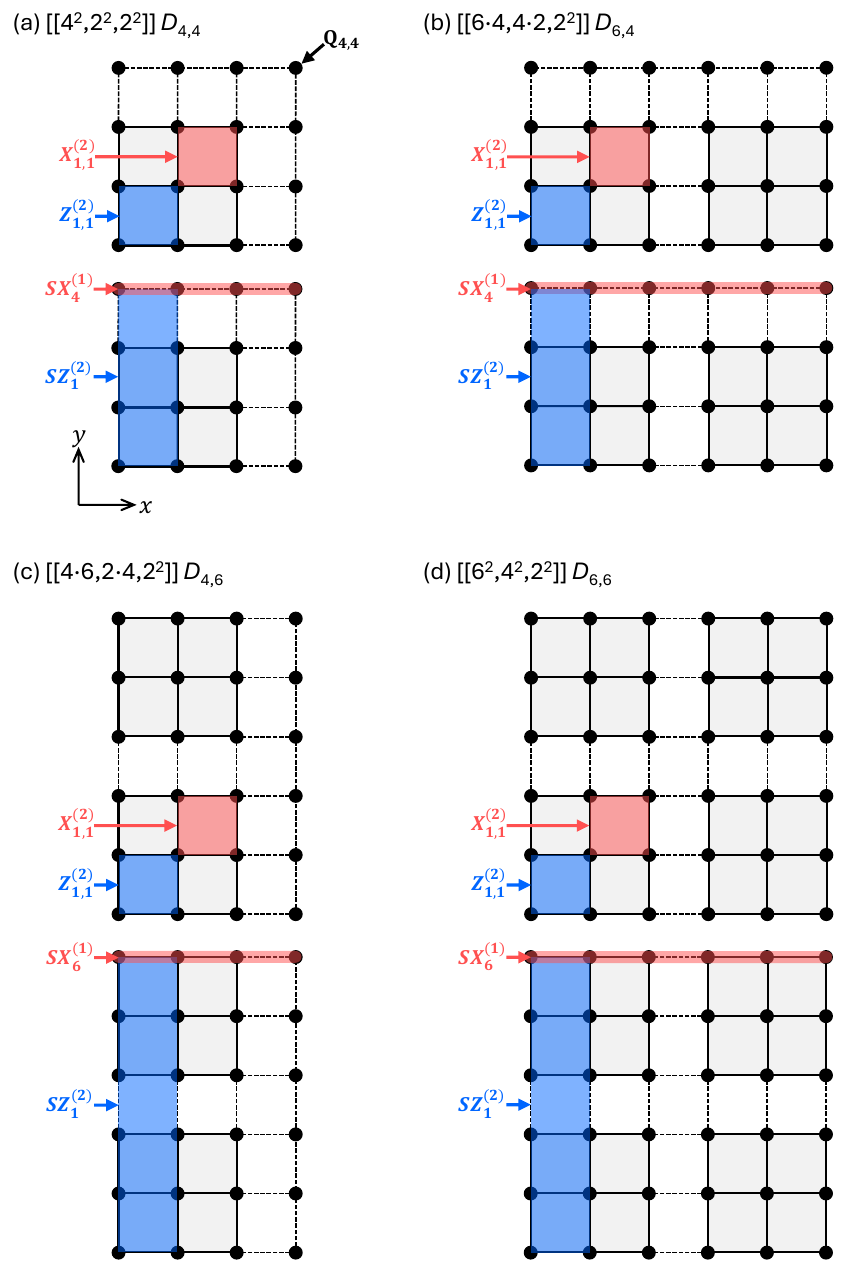}
    \caption{Visualization of level-2 MHC codes.
    The circles represent physical qubits.
    Gray squares correspond to logical Pauli operators.
    In each figure, the top shows logical Pauli operators for $\mathrm{Q}_{1,1}$ and the bottom shows examples of stabilizers 
    by highlighting them in blue and red for $Z$ and $X$ operators, 
    respectively.}
    \label{fig2}
\end{figure*}

A level-2 MHC code $D_{4,4}$ is 
obtained by concatenating $D_4$ with itself as follows.
We first prepare four level-1 code blocks encoded with $D_4$.
Referring to $Z_{i}^{(1)}$ and $X_{i}^{(1)}$ 
of the $j$th block as $Z_{i,j}^{(1)}$ and $X_{i,j}^{(1)}$, respectively, 
the logical Pauli operators of $D_{4,4}$ are defined as
\begin{align}
Z_{i,1}^{(2)}&=Z_{i,1}^{(1)} Z_{i,2}^{(1)}, 
X_{i,1}^{(2)}=X_{i,2}^{(1)} X_{i,3}^{(1)}, \nonumber\\
Z_{i,2}^{(2)}&=Z_{i,2}^{(1)} Z_{i,3}^{(1)}, 
X_{i,2}^{(2)}=X_{i,1}^{(1)} X_{i,2}^{(1)}. \nonumber
\end{align}
Note that all of them have weight of ${2^2=4}$, 
leading to code distance of 4.
Also, there are ${2^2=4}$ logical qubits.
Thus, $D_{4,4}$ is a $[[16,4,4]]$ code.
In addition to the original stabilizers of $D_4$, 
the following stabilizers are also introduced: 
\begin{align}
SZ_i^{(2)}&=Z_{i,1}^{(1)} Z_{i,2}^{(1)} Z_{i,3}^{(1)} Z_{i,4}^{(1)},
SX_i^{(2)}=X_{i,1}^{(1)} X_{i,2}^{(1)} X_{i,3}^{(1)} X_{i,4}^{(1)}. \nonumber
\end{align}
By stacking the four level-1 code blocks along the virtual $y$ axis, 
where $\mathrm{Q}_i$ of the $j$th block, 
which is denoted by $\mathrm{Q}_{i,j}$, is placed at ${(x,y)=(i,j)}$, 
the above operators can be visualized as shown in Fig.~\ref{fig2}(a).
Note that each logical Pauli operator corresponds to a square 
(two-dimensional hypercube).

Similarly, another level-2 MHC code $D_{6,4}$ 
obtained by using $D_6$ at level 1 and $D_4$ at level 2 
is defined by the same equations 
as $D_{4,4}$ except for ${i=1,\ldots,6}$ in the case of $D_{6,4}$.
$D_{6,4}$ is a $[[6\cdot 4,4\cdot 2,2^2]]=[[24,8,4]]$ code.
The operators for $D_{6,4}$ are visualized in Fig.~\ref{fig2}(b).

Another level-2 MHC code $D_{4,6}$ 
obtained by using $D_4$ at level 1 and $D_6$ at level 2 
is defined as follows:
\begin{align}
Z_{i,1}^{(2)}&=Z_{i,1}^{(1)} Z_{i,2}^{(1)}, 
X_{i,1}^{(2)}=X_{i,2}^{(1)} X_{i,3}^{(1)}, \nonumber \\
Z_{i,2}^{(2)}&=Z_{i,2}^{(1)} Z_{i,3}^{(1)}, 
X_{i,2}^{(2)}=X_{i,1}^{(1)} X_{i,2}^{(1)}, \nonumber \\
Z_{i,3}^{(2)}&=Z_{i,4}^{(1)} Z_{i,5}^{(1)}, 
X_{i,3}^{(2)}=X_{i,5}^{(1)} X_{i,6}^{(1)}, \nonumber \\
Z_{i,4}^{(2)}&=Z_{i,5}^{(1)} Z_{i,6}^{(1)}, 
X_{i,4}^{(2)}=X_{i,4}^{(1)} X_{i,5}^{(1)}, \nonumber \\
SZ_i^{(2)}&=Z_{i,1}^{(1)} Z_{i,2}^{(1)} Z_{i,3}^{(1)} Z_{i,4}^{(1)} Z_{i,5}^{(1)} Z_{i,6}^{(1)},& & \nonumber \\
SX_i^{(2)}&=X_{i,1}^{(1)} X_{i,2}^{(1)} X_{i,3}^{(1)} X_{i,4}^{(1)} X_{i,5}^{(1)} X_{i,6}^{(1)}.& &  \nonumber
\end{align}
$D_{6,6}$ is defined similarly.
$D_{4,6}$ and $D_{6,6}$ are 
$[[4\cdot 6,2\cdot 4,2^2]]=[[24,8,4]]$ and 
$[[6^2,4^2,2^2]]=[[36,16,4]]$ codes, respectively. 
These operators are visualized in 
Figs.~\ref{fig2}(c) and \ref{fig2}(d), respectively.

\subsection{Level-3 MHC codes}

\begin{figure*}[t]
    \includegraphics[width=1.6\columnwidth]{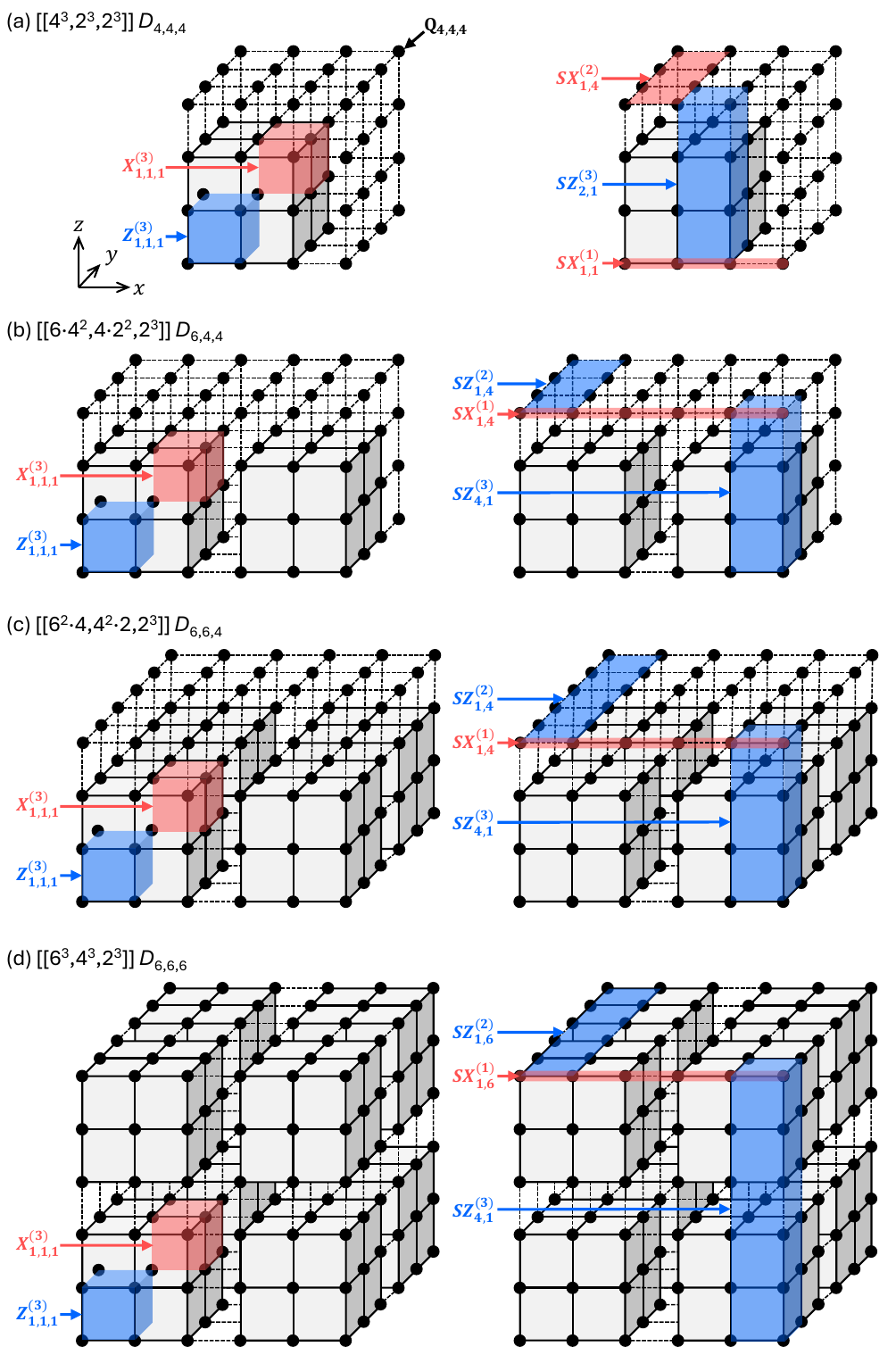}
    \caption{Visualization of level-3 MHC codes.
    The circles represent physical qubits.
    Gray cubes correspond to logical Pauli operators.
    In each figure, the left shows logical Pauli operators for $\mathrm{Q}_{1,1,1}$ and the right shows examples of stabilizers 
    by highlighting them in blue and red for $Z$ and $X$ operators, 
    respectively.}
    \label{fig3}
\end{figure*}

A level-3 MHC code $D_{4,4,4}$ using four $D_{4,4}$ blocks 
is defined as follows:
\begin{align}
Z_{i,j,1}^{(3)}&=Z_{i,j,1}^{(2)} Z_{i,j,2}^{(2)}, 
X_{i,j,1}^{(3)}=X_{i,j,2}^{(2)} X_{i,j,3}^{(2)}, \nonumber \\
Z_{i,j,2}^{(3)}&=Z_{i,j,2}^{(2)} Z_{i,j,3}^{(2)}, 
X_{i,j,2}^{(3)}=X_{i,j,1}^{(2)} X_{i,j,2}^{(2)}. \nonumber \\
SZ_{i,j}^{(3)}&=Z_{i,j,1}^{(2)} Z_{i,j,2}^{(2)} Z_{i,j,3}^{(2)} Z_{i,j,4}^{(2)}, \nonumber \\
SX_{i,j}^{(3)}&=X_{i,j,1}^{(2)} X_{i,j,2}^{(2)} X_{i,j,3}^{(2)} X_{i,j,4}^{(2)}, \nonumber
\end{align}
where $Z_{i,j,k}^{(2)}$ and  $X_{i,j,k}^{(2)}$ denote 
$Z_{i,j}^{(2)}$ and $X_{i,j}^{(2)}$, respectively, 
of the $k$th $D_{4,4}$ block.
$D_{4,4,4}$ also has the original stabilizers of $D_{4,4}$.
By stacking the four $D_{4,4}$ blocks along the virtual $z$ axis, 
where $\mathrm{Q}_{i,j}$ of the $k$th block, 
which is denoted by $\mathrm{Q}_{i,j,k}$, is placed at ${(x,y,z)=(i,j,k)}$, 
the above operators can be visualized as shown in Fig.~\ref{fig3}(a).
Each logical Pauli operator corresponds to a cube 
(three-dimensional hypercube).
The other level-3 MHC codes $D_{n_1,n_2,4}$ are defined similarly.
The code structures of $D_{6,4,4}$ and $D_{6,6,4}$ are visualized 
in Figs.~\ref{fig3}(b) and \ref{fig3}(c), respectively.

Another level-3 MHC code $D_{6,6,6}$  
obtained by concatenating $D_6$ using six $D_{6,6}$ blocks 
is defined as follows:
\begin{align}
Z_{i,j,1}^{(3)}&=Z_{i,j,1}^{(2)} Z_{i,j,2}^{(2)}, 
X_{i,j,1}^{(3)}=X_{i,j,2}^{(2)} X_{i,j,3}^{(2)}, \nonumber \\
Z_{i,j,2}^{(3)}&=Z_{i,j,2}^{(2)} Z_{i,j,3}^{(2)}, 
X_{i,j,2}^{(3)}=X_{i,j,1}^{(2)} X_{i,j,2}^{(2)}. \nonumber \\
Z_{i,j,3}^{(3)}&=Z_{i,j,4}^{(2)} Z_{i,j,5}^{(2)}, 
X_{i,j,3}^{(3)}=X_{i,j,5}^{(2)} X_{i,j,6}^{(2)}. \nonumber \\
Z_{i,j,4}^{(3)}&=Z_{i,j,5}^{(2)} Z_{i,j,6}^{(2)}, 
X_{i,j,4}^{(3)}=X_{i,j,4}^{(2)} X_{i,j,5}^{(2)}. \nonumber \\
SZ_{i,j}^{(3)}&=Z_{i,j,1}^{(2)} Z_{i,j,2}^{(2)} Z_{i,j,3}^{(2)} Z_{i,j,4}^{(2)} Z_{i,j,5}^{(2)} Z_{i,j,6}^{(2)}, \nonumber \\
SX_{i,j}^{(3)}&=X_{i,j,1}^{(2)} X_{i,j,2}^{(2)} X_{i,j,3}^{(2)} X_{i,j,4}^{(2)} X_{i,j,5}^{(2)} X_{i,j,6}^{(2)}. \nonumber
\end{align}
Some of these operators are shown in Fig.~\ref{fig3}(d).
The other level-3 MHC codes $D_{n_1,n_2,6}$ are defined similarly.

\subsection{Level-4 MHC codes}

The level-4 MHC codes $D_{n_1,n_2,n_3,4}$ and $D_{n_1,n_2,n_3,6}$ 
 are defined by similar equations 
to those for $D_{n_1,n_2,4}$ and $D_{n_1,n_2,6}$, respectively, 
in the last section. 
By introducing the fourth virtual axis denoted by $t$ and 
stacking level-3 $D_{n_1,n_2,n_3}$ blocks along the $t$ axis, 
each logical Pauli operator of the level-4 MHC codes 
corresponds to a four-dimensional hypercube, 
where $\mathrm{Q}_{i,j,k}$ of the $l$th $D_{n_1,n_2,n_3}$ block 
is placed at $(x,y,z,t)=(i,j,k,l)$.
This is the origin of the name of MHC codes.

\clearpage

\section{Performance comparison of MHC codes}
\label{sec3}

Here we numerically evaluate 
the performance of MHC codes for bit-flip errors 
without other errors, 
namely, code capacity. 
Note that because of the code symmetry with respect to $Z$ and $X$, 
the performance for bit-flip errors 
with error probability of $p_{\mathrm{flip}}$ 
is the same as that for depolarizing errors 
with error probability of $p_D=1.5p_{\mathrm{flip}}$. 
This is because in the depolarizing error model, 
one of the three Pauli errors, $X$, $Z$, and ${Y=iXZ}$, occurs 
with the same probability $p_D/3$ and therefore 
a bit-flip ($X$) error occurs with probability of $2p_D/3$.

In this numerical simulation, 
we first prepare an error-free all-zero state encoded with a MHC code.
The all-zero state encoded with $D_4$ is a four-qubit Greenberger-Home-Zeilinger (GHZ) state, 
\begin{align}
|00\rangle_{\mathrm{L}1} = \frac{|0000\rangle + |1111\rangle}{\sqrt{2}},
\end{align}
and that with $D_6$ is a six-qubit GHZ state, 
\begin{align}
|0000\rangle_{\mathrm{L}1} = \frac{|000000\rangle + |111111\rangle}{\sqrt{2}}.
\end{align}
Thus, the initial state can be prepared using the encoders 
in Figs.~\ref{fig5}(a) and \ref{fig5}(b), respectively, 
where we assume no errors on all the operations.
Next, we induce bit-flip errors on the physical qubits, 
then measure all the physical qubits in the $Z$ basis, 
and finally decode the measurement results using 
the level-by-level minimum-distance decoder 
developed in the original work~\cite{Goto2024a,comment_decoder}.
Unless all the logical-qubit values obtained by the decoding are 0, 
we consider that the decoding fails, 
leading to the block error probability.
In this work, we used a Python package called Stim~\cite{Gidney2021a} 
for quantum circuit simulations.

\begin{figure}[t]
    \includegraphics[width=0.6\columnwidth]{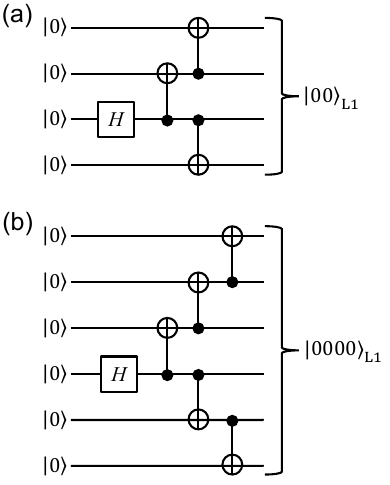}
    \caption{All-zero state encoders for (a) $D_4$ and (b) $D_6$.}
    \label{fig4}
\end{figure}

Figure~\ref{fig5}(a) shows the results for the four level-2 MHC codes.
The results are natural in the sense that lower-rate, smaller-size codes 
can achieve higher performance (lower decoding error probabilities) 
than higher-rate, larger-size codes.
On the other hand, 
it is noteworthy that $D_{6,4}$ outperforms $D_{4,6}$.
This is notable because it has been assumed that 
smaller-size codes should be used at lower levels~\cite{Liu2025a}.

We also evaluated the performance of level-3 and level-4 MHC codes 
at ${p_{\mathrm{flip}}=1\%}$ and 2\%, respectively, 
the results of which are shown in Figs.~\ref{fig5}(b) and \ref{fig5}(c), 
respectively. 
As in the case of level 2, 
our results clearly suggest that 
we should use larger-size codes for lower levels for MHC codes, 
contrary to the previous assumption~\cite{Liu2025a}.
More importantly, Figs.~\ref{fig5}(b) and \ref{fig5}(c) 
provide another counterintuitive result that 
the highest-performance MHC code is not the smallest-size one. 
The best MHC code is $D_{6,4,4}$ at level 3 and $D_{6,6,4,4}$ at level 4.
(The level-4 MHC code chosen in a previous work~\cite{Liu2025a}, 
$D_{4,4,6,6}$, is the worst.)
This is the main result of this work.

\begin{figure*}[t]
    \includegraphics[width=1.7\columnwidth]{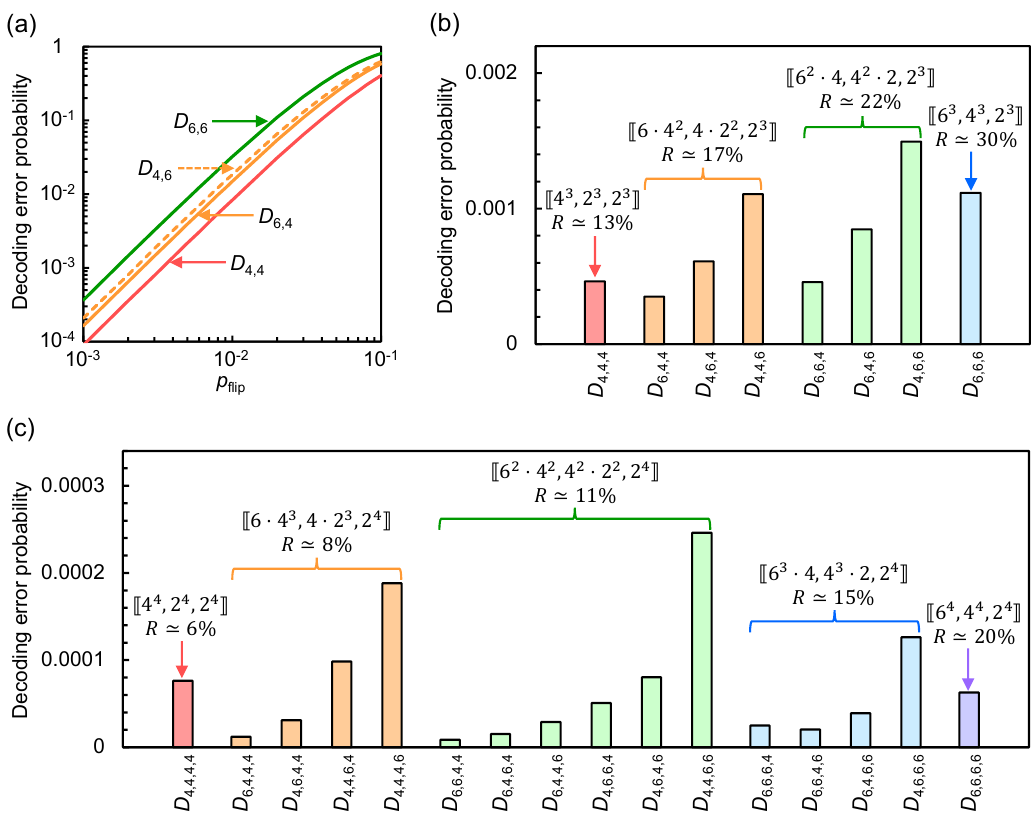}
    \caption{Performance comparison of MHC codes.
    (a--c) Decoding error probabilities at level 2--4, respectively, 
    for bit-flip errors.}
    \label{fig5}
\end{figure*}

\section{Proposed encoders of level-3 MHC codes}
\label{sec4}

Before performance evaluation in a circuit-level noise model, 
here we propose new fault-tolerant encoders for the level-3 MHC codes.

At level 1, we use the standard method for GHZ states 
with an ancilla physical qubit, 
as in the original work~\cite{Goto2024a}.
Thus, the total number of physical qubits necessary for 
preparing a $D_{n_1}$ all-zero state is 
$N^{\prime (1)}_{n_1}=n_1+1$.

\begin{figure*}[t]
    \includegraphics[width=1.6\columnwidth]{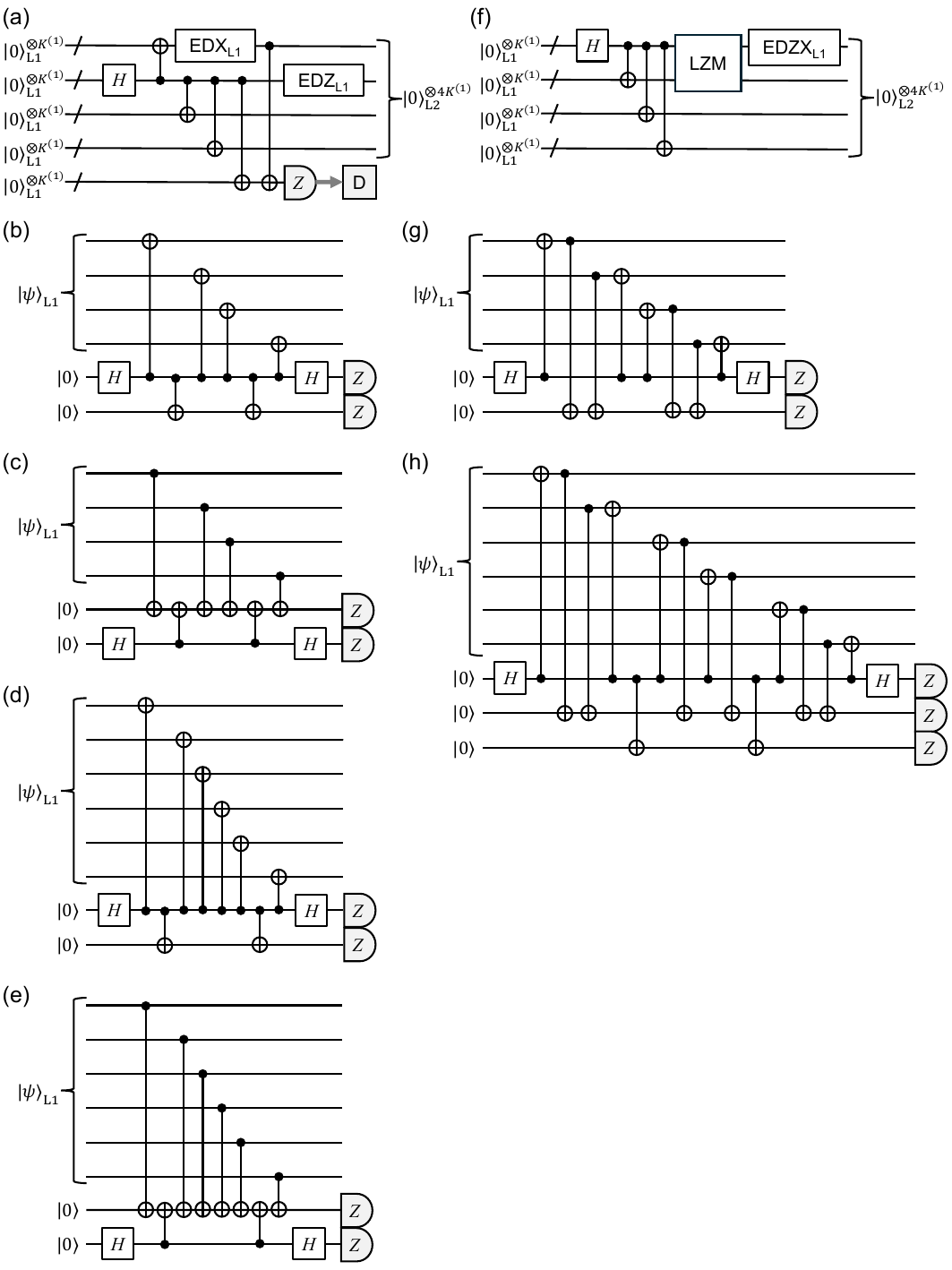}
    \caption{Fault-tolerant encoders for level-2 MHC codes. 
    Here we show the encoders for only $D_{n,4}$ (${n=4, 6}$), 
    but $D_{n,6}$ encoders are similar.
    $K^{(1)}$ denotes the number of logical qubits in the level-1 block.
    (a) Original encoder. EDZ$_{\mathrm{L}1}$ and EDX$_{\mathrm{L}1}$ are 
    level-1 $Z$-error and $X$-error detection gadgets, respectively, 
    which are implemented as in (b) and (c), respectively, for $D_4$ and 
    as in (d) and (e), respectively, for $D_6$. 
    (b,c) $Z$-error (b) and $X$-error (c) detection for $D_4$ 
    with a flag qubit.
    (d,e) $Z$-error (d) and $X$-error (e) detection for $D_6$ 
    with a flag qubit.
    (f) Proposed encoder. 
    LZM is a logical $Z$ operator 
    measurement gadget implemented 
    as in Fig.~\ref{fig7}. 
    EDZX$_{\mathrm{L}1}$ is 
    a level-1 simultaneous $Z$-error and $X$-error detection gadget 
    implemented as in (g) for $D_4$ and 
    as in (h) for $D_6$.
    (g,h) Simultaneous $Z$-error and $X$-error detection for $D_4$ (g) 
    and $D_6$ (h).}
    \label{fig6}
\end{figure*}

\begin{figure*}[t]
    \includegraphics[width=1.2\columnwidth]{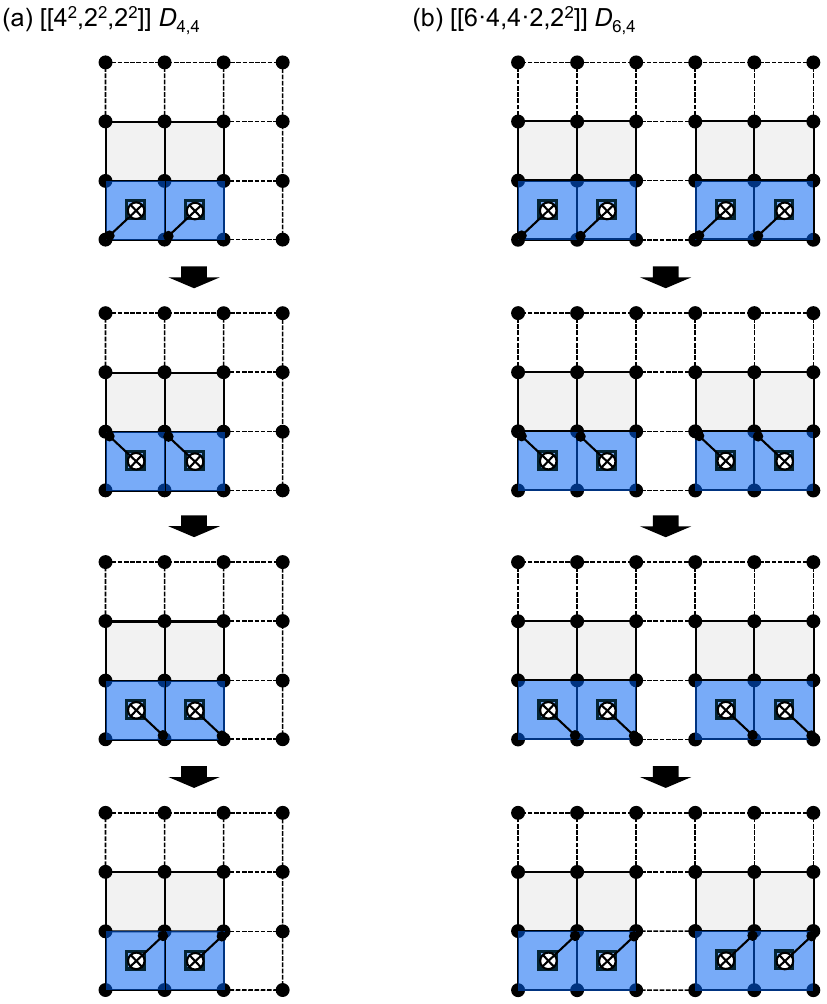}
    \caption{Direct measurements of level-2 logical $Z$ operators. 
    Here we show only the cases of $D_{n,4}$ (${n=4, 6}$), 
    but the cases of $D_{n,6}$ are similar. 
    The small squares are physical ancilla qubits used for the measurements, 
    which are initially set in $|0\rangle$.
    (a,b) Gate sequence for the logical $Z$ measurements used 
    in the $D_{4,4}$ (a) and $D_{6,4}$ (b) encoders.
    Note that the bottom two level-1 blocks correspond to 
    the top two ones in Fig.~\ref{fig6}(f).}
    \label{fig7}
\end{figure*}

At level 2, the encoder in Fig.~\ref{fig6}(a) was proposed 
in the original work~\cite{Goto2024a}, 
where the error-detection gadgets are implemented 
by the flag method~\cite{Chao2018a} 
as in Figs.~\ref{fig6}(b) and \ref{fig6}(c) for $D_4$ and 
as in Figs.~\ref{fig6}(d) and \ref{fig6}(e) for $D_6$, 
and also logical $X$ errors are removed by using a logical ancilla block.
The proposed encoder is shown in Fig.~\ref{fig6}(f), 
where the $Z$-error and $X$-error detection is performed at once more efficiently by the method in Figs.~\ref{fig6}(g) for $D_4$ 
or \ref{fig6}(h) for $D_6$, 
and also we do not use a logical ancilla block.
The simultaneous measurements of weight-4 $Z$ and $X$ stabilizers 
in Fig.~\ref{fig6}(g) were proposed in Ref.~\citenum{Reichardt2018a}, 
where the ancilla qubits play both the roles of error detection and flag, 
and therefore additional flag qubits are unnecessary. 
That is, this is more efficient 
than the methods in Figs.~\ref{fig6}(b) and \ref{fig6}(c).
The simultaneous measurements of weight-6 $Z$ and $X$ stabilizers 
in Fig.~\ref{fig6}(h) are based on 
the proposal in Ref.~\citenum{Self2024a}, 
but we modify it by adding a flag qubit to remove correlated $X$ errors, 
leading to higher performance.
To remove logical $X$ errors, instead of using a logical ancilla block, 
we directly measure logical $Z$ operators of the level-2 MHC codes 
using physical ancilla qubits, as shown in Fig.~\ref{fig7}.
Crucial correlated $Z$ errors caused by these measurements 
can be removed by the following error detection, 
making the encoder fault-tolerant.
To achieve this, 
the order of the physical CNOT gates in Fig.~\ref{fig7} is important.

The total number of physical qubits necessary for 
preparing a $D_{n_1,n_2}$ all-zero state is 
\begin{align}
N^{\prime (2)}_{n_1,n_2}
&=(n_2+1) N^{(1)}_{n_1} + 2 + 2
\nonumber \\
&= N^{(2)}_{n_1,n_2} + n_2 + n_1 + 5
\end{align}
for the original encoder and 
\begin{align}
N^{\prime (2)}_{n_1,n_2}
&=n_2 N^{(1)}_{n_1} + (n_1-2) + n_1/2 
\nonumber \\
&= N^{(2)}_{n_1,n_2} + n_2 + 3n_1/2 -2
\end{align}
for the proposed encoder.
(Note that the latter formula holds only for ${n_1, n_2 = 4, 6}$.)
Here, $N^{(2)}_{n_1,n_2}=n_2 n_1$ on the right-hand side 
is the number of physical qubits in the level-2 code block, 
and the other terms are the number of ancilla qubits, namely, overheads.
Thus, the proposed encoder can reduce the overhead by 
$7-n_1/2$.

\begin{figure*}[t]
    \includegraphics[width=1.6\columnwidth]{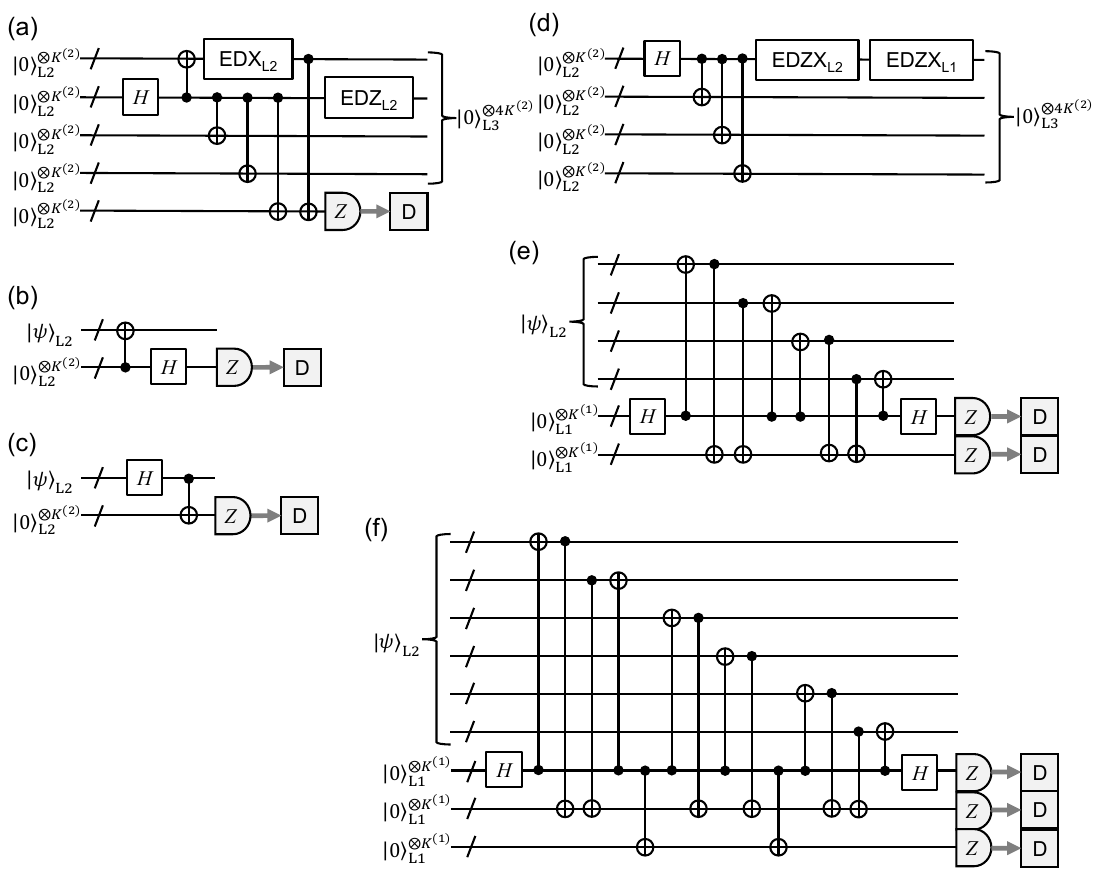}
    \caption{Fault-tolerant encoders for level-3 MHC codes. 
    Here we show the encoders for only $D_{n_1,n_2,4}$ (${n_1, n_2 =4, 6}$), 
    but $D_{n_1,n_2,6}$ encoders are similar.
    $K^{(L)}$ ({$L=1, 2$}) denotes the number of logical qubits 
    in the level-$L$ block.
    (a) Original encoder. EDZ$_{\mathrm{L}2}$ and EDX$_{\mathrm{L}2}$ are 
    level-2 $Z$-error and $X$-error detection gadgets, respectively, 
    which are implemented as in (b) and (c), respectively. 
    (b,c) $Z$-error (b) and $X$-error (c) detection based on the Steane method.
    (d) Proposed encoder. 
    EDZX$_{\mathrm{L}2}$ is 
    a level-2 simultaneous $Z$-error and $X$-error detection gadget implemented as in (e) for $D_{n_1,4}$ and 
    as in (f) for $D_{n_1,6}$ ({$n_1=4, 6$}).
    EDZX$_{\mathrm{L}1}$ is 
    level-1 simultaneous $Z$-error and $X$-error detection gadgets 
    implemented as in Fig.~\ref{fig6}(g) for $D_4$ and 
    as in Fig.~\ref{fig6}(h) for $D_6$.
    (e,f) Simultaneous $Z$-error and $X$-error detection for $D_{n_1,4}$ (e) 
    and $D_{n_1,6}$ (f) ({$n_1=4, 6$}).}
    \label{fig8}
\end{figure*}

The original level-3 encoder shown in Fig.~\ref{fig8}(a) is 
similar to the level-2 one in Fig.~\ref{fig6}(a), 
but different implementations based on the Steane method~\cite{Steane1997a} 
[Figs.~\ref{fig8}(b) and \ref{fig8}(c)] are 
adopted for the error-detection gadgets~\cite{Goto2024a}.
The proposed encoder shown in Fig.~\ref{fig8}(d) 
is similar to the level-2 one in Fig.~\ref{fig6}(f), 
but it does not measure logical $Z$ operators. 
This is enough if we use the level-3 logical qubits for quantum computation
(not use for level-4 logical state preparation).
Also, we perform simultaneous $Z$-error and $X$-error 
detection for not only level-1 blocks using the method 
in Fig.~\ref{fig6}(g) or \ref{fig6}(h), but also 
a level-2 block using the method in  Fig.~\ref{fig8}(e) or \ref{fig8}(f).

The total number of physical qubits necessary for 
preparing a $D_{n_1,n_2,n_3}$ all-zero state is 
\begin{align}
N^{\prime (3)}_{n_1,n_2,n_3}
&=(n_3+1) N^{\prime (2)}_{n_1,n_2} + N^{\prime (2)}_{n_1,n_2} 
+ N^{\prime (2)}_{n_1,n_2}
\nonumber \\
&= N^{(3)}_{n_1,n_2,n_3}+ n_3 n_2 + n_3 n_1 + 3n_2 n_1
\nonumber \\
&+ 5n_3+ 3n_2 + 3n_1 + 15
\end{align}
for the original encoder and 
\begin{align}
N^{\prime (3)}_{n_1,n_2,n_3}
&=n_3 N^{\prime (2)}_{n_1,n_2} + (n_2/2) N^{\prime (1)}_{n_1} + n_2 (n_1/2) 
\nonumber \\
&= N^{(3)}_{n_1,n_2,n_3} + n_3 n_2 + 3 n_3 n_1/2 + n_2 n_1 
\nonumber \\
&-2n_3  + n_2/2
\end{align}
for the proposed encoder, 
where $N^{(3)}_{n_1,n_2,n_3} = n_1 n_2 n_3$.
(Note that the latter formula holds only for $n_1, n_2, n_3 = 4, 6$.)
Thus, the proposed encoder can reduce the overhead by 
$2n_2 n_1 + 7n_3 + 5n_2/2 + 3n_1 + 15- n_3 n_1/2$.
This is about 60\% of the original overhead 
for all the eight level-3 MHC codes.
Thus, the proposed encoder realizes about 60\% reduction of overhead.

\clearpage

\section{Logical CNOT performance}
\label{sec5}

Using the encoders presented in the last section, 
we evaluate the performance of 
four promising level-3 MHC codes, namely, 
$D_{4,4,4}$, $D_{6,4,4}$, $D_{6,6,4}$, and $D_{6,6,6}$, 
in a circuit-level noise model. 
In the present error model 
with error rate $p_{\mathrm{circ}}$, 
we assume that 
each of physical-qubit zero-state preparation and $Z$-basis measurement is accompanied by a bit flip with probability $p_{\mathrm{circ}}$, 
each physical CNOT gate is followed by one of 15 two-qubit Pauli errors 
with equal probability $p_{\mathrm{circ}}/15$, 
and there are no single-qubit-gate and memory errors, 
as in the original work~\cite{Goto2024a}. 
This model may be relevant for 
ion-trap~\cite{Pino2021a,Ryan2021a,Postler2022a,Moses2023a,Wang2024a,Postler2024a,Self2024a,Yamamoto2024a,Silva2024a,Ryan2024a,Dasu2025a,Daguerre2025a,Hughes2025a,Ransford2025a} and 
neutral-atom~\cite{Graham2022a,Bluvstein2022a,Evered2023a,Bluvstein2024a,Muniz2025a,Bluvstein2025a,Rodriguez2025a,Chiu2025a,Manetsch2025a,Infleqtion2025a} systems. 

To evaluate the performance 
in the circuit-level noise model, 
we numerically simulate 
logical-CNOT gates, 
which allows for performance evaluation against not only 
$X$ errors but also $Z$ errors. 
(If we evaluate logical zero states generated by the proposed encodes, 
then we can evaluate 
the performance 
only against $X$ errors.)
The logical-CNOT error probability is estimated as follows~\cite{Goto2024a}.
First, two error-free logical Bell pairs are prepared 
by performing error-free logical Hadamard and CNOT gates on 
error-free logical all-zero states.
Next, we perform ten times 
transversal physical CNOT gates followed 
by error-correcting teleportation~\cite{Knill2005a,Goto2024a,Knill2005b} 
on the first blocks of the two Bell-pair blocks, 
where we use the encoders in Sec.~\ref{sec4} and 
physical operations are faulty according to the above error model.
Then, we disentangle the logical Bell pairs 
by performing error-free logical Hadamard and CNOT gates. 
Last, we measure all the physical qubits in the $Z$ basis 
and decode the results. 
We define the block error probability  
as ${p_{\mathrm{block}}=1-(1-p_{10})^{1/10}}$, 
where $p_{10}$ is the probability that 
at least one of the logical-qubit values obtained by the decoding is 1.
The error probability per logical CNOT, $p_{\mathrm{CNOT}}$, 
is also defined as $p_{\mathrm{CNOT}}=1-(1-p_{\mathrm{block}})^{1/K^{(3)}}$, 
where $K^{(3)}$ is the number of logical qubits in the level-3 block.

\begin{figure*}[t]
    \includegraphics[width=1.5\columnwidth]{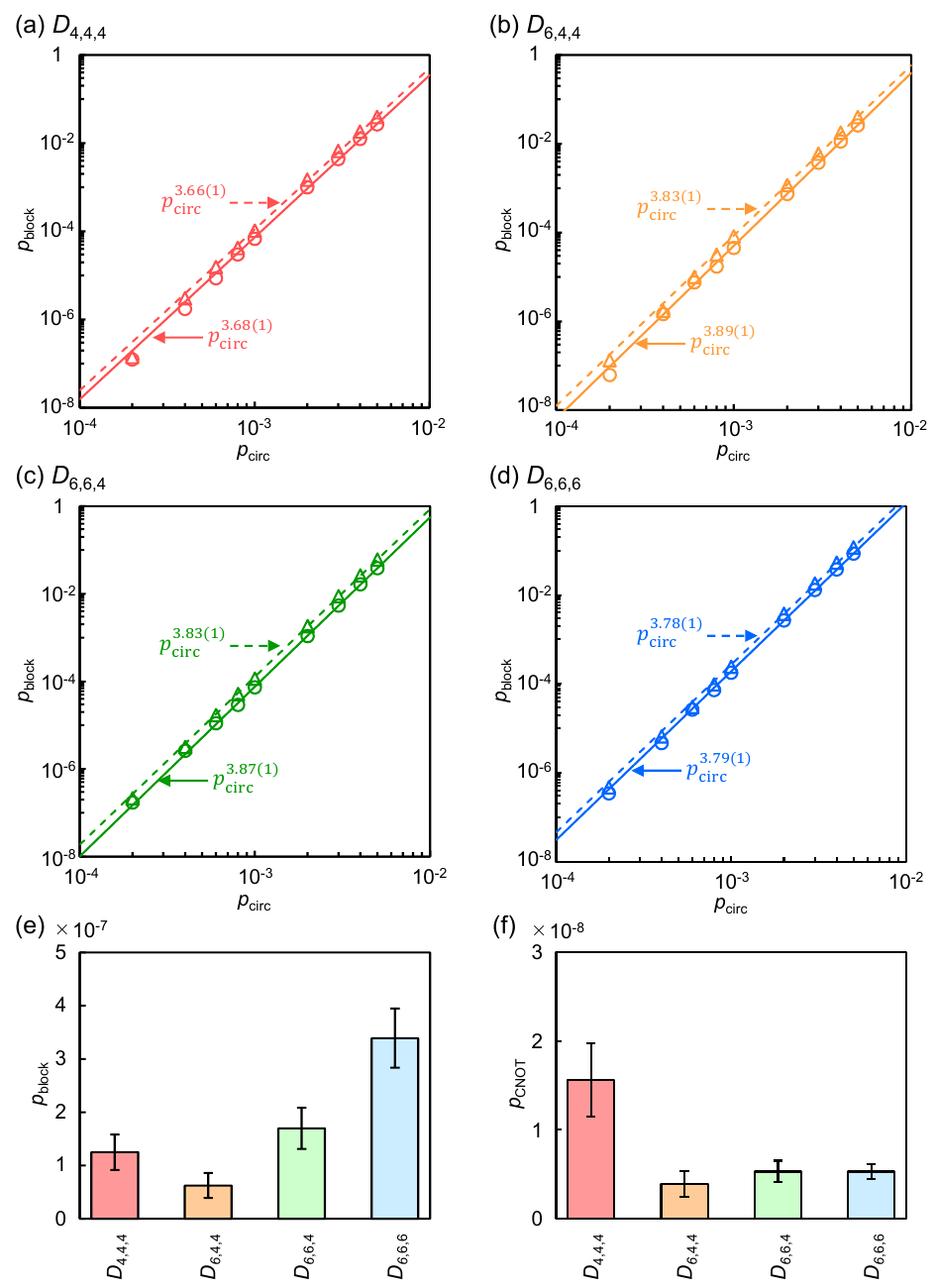}
    \caption{Numerical results of logical-CNOT error probability 
    for level-3 MHC codes. 
    (a--d) Log-log plots of the block error probability $p_{\mathrm{block}}$ for $D_{4,4,4}$, $D_{6,4,4}$, $D_{6,6,4}$, and $D_{6,6,6}$, 
    respectively.
    $p_{\mathrm{block}}$ is the probability that at least one logical CNOT fails per block
    (see the main text for its definition).
    The circles and triangles represent the results with 
    the proposed and original encoders, respectively.
    The solid and dashed lines are the fitting results 
    with the power function $\beta p_{\mathrm{circ}}^{\alpha}$ 
    ($\alpha$ and $\beta$ are fitting parameters)
    to the circles and triangles, respectively.
    In each panel, $p_{\mathrm{circ}}^{\alpha}$ shows 
    the fitting result of $\alpha$.
    (e) $p_{\mathrm{block}}$ at ${p_{\mathrm{circ}}=2\times 10^{-4}}$.
    (f) Logical-CNOT error probability $p_{\mathrm{CNOT}}$ 
    at ${p_{\mathrm{circ}}=2\times 10^{-4}}$.
    $p_{\mathrm{CNOT}}$ is defined 
    as $p_{\mathrm{CNOT}}=1-(1-p_{\mathrm{block}})^{1/K^{(3)}}$, 
    where $K^{(3)}$ is the number of logical qubits in the level-3 block.}
    \label{fig9}
\end{figure*}

Figure~\ref{fig9} shows the results for the four level-3 MHC codes 
with the original and proposed encoders.
All the exponents obtained by the power-function fitting 
are close to ${d/2=4}$ (${d=8}$ is the code distance).
This suggests the fault tolerance of the original and proposed encoders.
Also, all the solid lines are a little lower 
than the corresponding dashed lines.
This means that the proposed encoders are a little better than 
the original ones.
Most importantly, 
$D_{6,4,4}$ achieves the lowest values for 
both $p_{\mathrm{block}}$ and $p_{\mathrm{CNOT}}$.
This concludes that $D_{6,4,4}$ is the best of the level-3 MHC codes.

\begin{figure*}[t]
    \includegraphics[width=1.5\columnwidth]{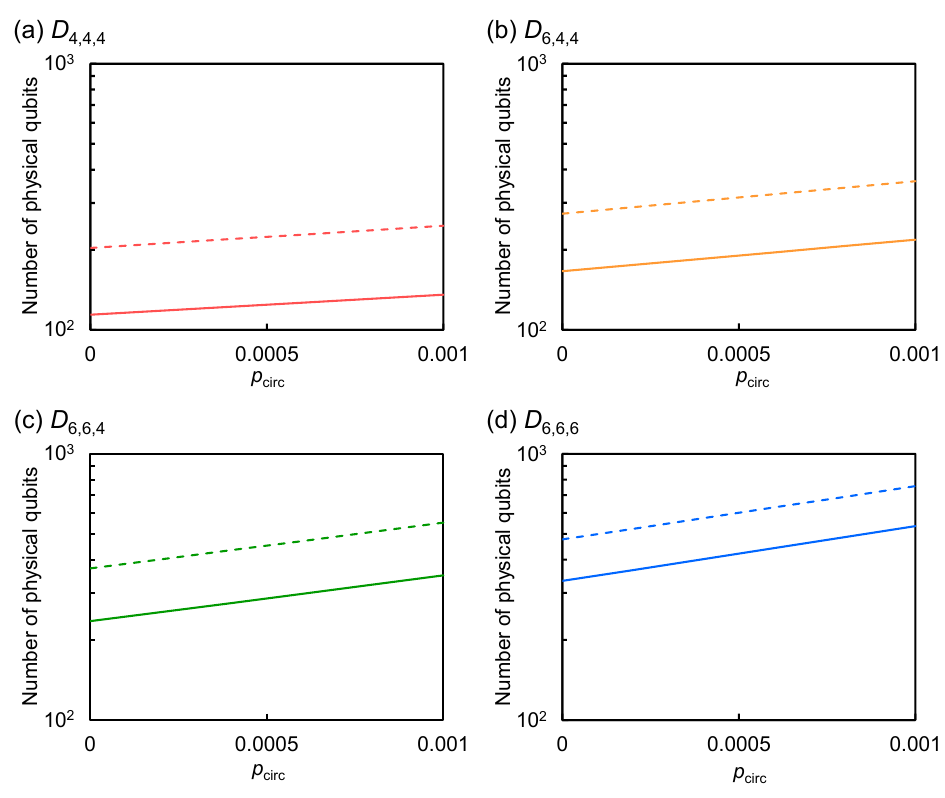}
    \caption{Total number of physical qubits 
    necessary for preparing a logical all-zero state for level-3 MHC codes. 
    (a--d) Semi-log plots of the total number of physical qubits 
    for $D_{4,4,4}$, $D_{6,4,4}$, $D_{6,6,4}$, and $D_{6,6,6}$, 
    respectively.
    The solid and dashed curves show the results for 
    the proposed and original encoders, respectively.}
    \label{fig10}
\end{figure*}

Figure~\ref{fig10} shows the total number of physical qubits 
necessary for preparing a logical all-zero state, 
taking the repetition due to error detection into account.
The results clearly show that 
the proposed encoders can reduces the overheads compared with 
the original encoders even when the error-detection effect is included.

\section{Summary and outlook}
\label{sec6}

Toward lower logical error rates and earlier experimental realization of 
MHC codes, 
we have comprehensively investigated the MHC codes 
with $[[6,4,2]]$ and/or $[[4,2,2]]$ quantum error-detecting codes.
Consequently, we have found a counterintuitive fact that 
$D_{6,4,4}$ and $D_{6,6,4,4}$ achieve the lowest decoding error probability 
among the level-3 and level-4 MHC codes, respectively.
Focusing on level 3, we have also developed efficient 
fault-tolerant encoders realizing about 60\% overhead reduction 
compared with the original encoders, while maintaining or even improving logical-CNOT performance.
It have turned out that 
$D_{6,4,4}$ achieves the highest performance for logical CNOT gates.
Thus, we conclude that $D_{6,4,4}$ is the best of the level-3 MHC codes.
This code can achieve not only lower logical error rates 
but also the smaller number of necessary physical qubits, 
compared to the original $D_{6,6,6}$. 
Therefore, $D_{6,4,4}$ will become a good target 
toward early experimental realization of 
efficient FTQC with a high-rate quantum code.

In this work, we have evaluated only logical CNOT gates and 
left the evaluation of other logical gates for future work.
Also, we have assumed no memory errors 
in the present circuit-level noise model.
Toward experimental realization, e.g., with trapped ions or neutral atoms, 
we should optimize the gate sequence 
to minimize the moving distances of qubits.
Such optimization is also left for future work.

\begin{acknowledgments}
This work was supported by JST Moonshot R\&D
Grant Number JPMJMS2061. 
Parts of numerical simulations were done on the HOKUSAI supercomputer
at RIKEN (project ID RB230022).
\end{acknowledgments}


\clearpage


\begin{thebibliography}{99}

\bibitem{Shor1995a}
P. W. Shor, 
Scheme for reducing decoherence in quantum computer memory, 
Phys. Rev. A \textbf{52}, R2493 (1995). 

\bibitem{Steane1996a}
A. M. Steane, 
Error correcting codes in quantum theory, 
Phys. Rev. Lett. \textbf{77}, 793 (1996).


\bibitem{Calderbank1996a}
A. R. Calderbank and P. W. Shor, 
Good quantum error-correcting codes exist, 
Phys. Rev. A \textbf{54}, 1098 (1996).


\bibitem{Steane1996b}
A. M. Steane, 
Multiple-particle interference and quantum error correction, 
Proc. Roy. Soc. London Ser. A \textbf{452}, 2551 (1996).

\bibitem{Steane1996c}
A. M. Steane, 
Simple quantum error-correcting codes, 
Phys. Rev. A \textbf{54}, 4741 (1996).


\bibitem{Laflamme1996a}
R. Laflamme, C. Miquel, J.-P. Paz, and W. H. Zurek, 
Perfect quantum error-correcting code, 
Phys. Rev. Lett. \textbf{77}, 198 (1996).



\bibitem{Gottesman1996a}
D. Gottesman, 
Class of quantum error-correcting codes saturating the quantum Hamming bound, 
Phys. Rev. A \textbf{54}, 1862 (1996).




\bibitem{Shor1996a}
P. W. Shor, 
Fault-tolerant quantum computation, 
in \textit{Proc. 37th Symposium on Foundations of Computer Science} 
(IEEE Computer Society Press, 1996), pp. 56-65.

\bibitem{Gottesman1998a}
D. Gottesman, 
Theory of fault-tolerant quantum computation, 
Phys. Rev. A \textbf{57}, 127 (1998).


\bibitem{Nielsen}
M. A. Nielsen and I. L. Chuang, 
\textit{Quantum Computation and Quantum Information} 
(Cambridge University Press, Cambridge, England, 2000).



\bibitem{Aharonov1997a}
D. Aharonov and M. Ben-Or, 
Fault-tolerant quantum computation with constant error, 
in \textit{Proc. 29th Annu. ACM Symp. Theory Comput.} 
(ACM, El Paso, Texas, USA, 1997), p. 176.


\bibitem{Knill1996a}
E. Knill and R. Laflamme, 
Concatenated quantum codes, 
arXiv:quant-ph/9608012.


\bibitem{Zalka1996a}
C. Zalka, 
Threshold Estimate for Fault Tolerant Quantum
Computing,
arXiv:quant-ph/9612028.



\bibitem{Knill1998a}
E. Knill, R. Laflamme, and W. H. Zurek, 
Resilient quantum computation, 
Nature \textbf{279}, 342 (1998).




\bibitem{Steane2003a}
A. M. Steane, 
Overhead and noise threshold of fault-tolerant quantum error correction, 
Phys. Rev. A \textbf{68}, 042322 (2003). 

\bibitem{Knill2005a}
E. Knill, Quantum computing with realistically noisy devices, Nature \textbf{434}, 39 (2005).


\bibitem{Aliferis2006a}
P. Aliferis, D. Gottesman, and J. Preskill, 
Quantum accuracy threshold for concatenated distance-3 code, 
Quant. Inf. Comput. \textbf{6}, 97 (2006).


\bibitem{Goto2009a}
H. Goto and K. Ichimura, 
Fault-tolerant quantum computation with probabilistic two-qubit gates,
Phys. Rev. A \textbf{80}, 040303 (2009).

\bibitem{Goto2010a}
H. Goto and K. Ichimura,
Condition for fault-tolerant quantum computation with a cavity-QED scheme,
Phys. Rev. A \textbf{82}, 032311 (2010).

\bibitem{Fujii2010a}
K. Fujii and K. Yamamoto,
Cluster-based architecture for fault-tolerant quantum computation, 
Phys. Rev. A \textbf{81}, 042324 (2010).



\bibitem{Goto2013a}
H. Goto and H. Uchikawa,
Fault-tolerant quantum computation with a soft-decision decoder for error correction and detection by teleportation,
Sci. Rep. \textbf{3}, 2044 (2013).


\bibitem{Goto2014a}
H. Goto,
Step-by-step magic state encoding for efficient fault-tolerant quantum computation,
Sci. Rep. \textbf{4}, 7501 (2014).

\bibitem{Goto2014b}
H. Goto and H. Uchikawa,
Soft-decision decoder for quantum erasure and probabilistic-gate error models,
Phys. Rev. A \textbf{89}, 022322 (2014).









\bibitem{Kitaev1996a}
A. Y. Kitaev, 
Fault-tolerant quantum computation by anyons, 
arXiv:quant-ph/9707021.

\bibitem{Bravyi1998a}
S. B. Bravyi and A. Y. Kitaev, 
Quantum codes on a lattice with boundary, 
quant-ph/9811052 (1998).


\bibitem{Dennis2002a}
E. Dennis, A. Kitaev, A. Landahl, and J. Preskill, 
Topological quantum memory, 
J. Math. Phys. (N.Y.) \textbf{43}, 4452 (2002).


\bibitem{Raussendorf2007a}
R. Raussendorf and J. Harrington, 
Fault-Tolerant Quantum Computation with High Threshold in Two Dimensions, 
Phys. Rev. Lett. \textbf{98}, 190504 (2007).

\bibitem{Raussendorf2007b}
R. Raussendorf, J. Harrington, and K. Goyal, 
Topological fault-tolerance in cluster state quantum computation, 
New J. Phys. \textbf{9}, 199 (2007).

\bibitem{Fowler2009a}
A. G. Fowler, A. M. Stephens, and P. Groszkowski, 
High-threshold universal quantum computation on the surface code, 
Phys. Rev. A \textbf{80}, 052312 (2009).

\bibitem{Wang2011a}
D. S. Wang, A. G. Fowler, and L. C. L. Hollenberg, 
Surface code quantum computing with error rates over 1\%, 
Phys. Rev. A \textbf{83} 020302(R) (2011). 


\bibitem{Fowler2012a}
A. G. Fowler, M. Mariantoni, J. M. Martinis, and A. N. Cleland, 
Surface codes: Towards practical large-scale quantum computation, 
Phys. Rev. A \textbf{86}, 032324 (2012).


\bibitem{Horsman2022a}
D. Horsman, A. G. Fowler, S. Devitt, and R. Van Meter, 
Surface code quantum computing by lattice surgery, 
New J. Phys. \textbf{14} 123011 (2012).



\bibitem{Krinner2022a}
S. Krinner, N. Lacroix, A. Remm, A. Di Paolo, E. Genois, C. Leroux \textit{et al}., 
Realizing repeated quantum error correction in a distance-three surface code, 
Nature \textbf{605}, 669 (2022).

\bibitem{Zhao2022a}
Y. Zhao, Y. Ye, H.-L. Huang, Y. Zhang, D. Wu, H. Guan \textit{et al}., 
Realization of an Error-Correcting Surface Code with Superconducting Qubits, 
Phys. Rev. Lett. \textbf{129}, 030501 (2022).

\bibitem{Google2023a}
Google Quantum AI, 
Suppressing quantum errors by scaling a surface code logical qubit, 
Nature \textbf{614}, 676 (2023).

\bibitem{Google2024a}
Google Quantum AI and Collaborators, 
Quantum error correction below the surface code threshold,
Nature \textbf{638}, 920 (2025).




\bibitem{Breuckmann2021a}
N. P. Breuckmann and J. N. Eberhardt, 
Quantum Low-Density Parity-Check Codes, 
PRX Quantum \textbf{2}, 040101 (2021).

\bibitem{Gottesman2014a}
D. Gottesman, 
Fault-tolerant quantum computation with constant overhead, 
Quantum Info. Comput. \textbf{14}, 1338 (2014).

\bibitem{MacKay2004a}
D. J. C. MacKay, G. Mitchison, and P. L. McFadden, 
Sparse Graph Codes for Quantum Error-Correction, 
IEEE Transactions on Information Theory \textbf{50}, 2315 (2004).

\bibitem{Kovalev2013a}
A. A. Kovalev and L. P. Pryadko, 
Quantum Kronecker sum-product low-density parity-check codes with finite rate, 
Phys. Rev. A \textbf{88}, 012311 (2013).

\bibitem{Leverrier2015a}
A. Leverrier, J.-P. Tillich, and G. Z\'{e}mor, 
Quantum Expander Codes, 
in \textit{2015 IEEE 56th Annual Symposium on Foundations of Computer Science} 
(Berkeley, USA, 2015), pp. 810--824.

\bibitem{Fawzi2018a}
O. Fawzi, A. Grospellier, and A. Leverrier, 
Constant Overhead Quantum Fault-Tolerance with Quantum Expander Codes, 
Communications of the ACM \textbf{64}, 106 (2020).

\bibitem{Breuckmann2021a}
N. P. Breuckmann and J. N. Eberhardt, 
Balanced Product Quantum Codes, 
IEEE Transactions on Information Theory \textbf{67}, 6653 (2021).

\bibitem{Panteleev2022a}
P. Panteleev and G. Kalachev, 
Asymptotically Good Quantum  andLocally Testable Classical LDPC Codes, 
in \textit{Proc. 54th Annual ACMSIGACT Symposium on Theory of Computing} 
(STOC'22, Rome, Italy, 2022), pp. 375--388.

\bibitem{Dinur2023a}
I. Dinur, M.-H. Hsieh, T.-C. Lin, and T. Vidick, 
Good Quantum LDPC Codes with Linear Time Decoders, 
in \textit{Proc. 55th Annual ACM Symposium on Theory of Computing} 
(STOC'23, Orlando, USA, 2023), pp. 905--918.

\bibitem{Leverrier2022a}
A. Leverrier and G. Z\'{e}mor, 
Quantum Tanner codes, 
in \textit{2022 IEEE 63rd Annual Symposium on Foundations of Computer Science} 
(FOCS, Denver, USA, 2022), pp. 872--883.


\bibitem{Lin2022a}
T.-C. Lin and M.-H. Hsieh, 
Good quantum LDPC codes with linear time decoder from lossless expanders, 
arXiv:2203.03581.





\bibitem{Gottesman2022a}
D. Gottesman, 
Opportunities and Challenges in Fault-Tolerant Quantum Computation, 
arXiv:2210.15844 (2022).


\bibitem{Tremblay2022a}
M. A. Tremblay, N. Delfosse, and M. E. Beverland, 
Constant-Overhead Quantum Error Correction with Thin Planar Connectivity, 
Phys. Rev. Lett. \textbf{129}, 050504 (2022).

\bibitem{Viszlai2023a}
J. Viszlai, W. Yang, S. F. Lin, J. Liu, N. Nottingham, J. M. Baker, and F. T. Chong, 
Matching Generalized-Bicycle Codes to Neutral Atoms for Low-Overhead Fault-Tolerance, 
arXiv:2311.16980.


\bibitem{Bravyi2024a}
S. Bravyi, A. W. Cross, J. M. Gambetta, D. Maslov, P. Rall, and T. J. Yoder, 
High-threshold and low-overhead fault-tolerant quantum memory, 
Nature \textbf{627}, 778 (2024).

\bibitem{Xu2024a}
Q. Xu, J. P. B. Ataides, C. A. Pattison, N. Raveendran, D. Bluvstein, J. Wurtz, 
B. Vasi\'{c}, M. D. Lukin, L. Jiang, and H. Zhou, 
Constant-Overhead Fault-Tolerant Quantum Computation with Reconfigurable Atom Arrays, 
Nat. Phys. \textbf{20}, 1084 (2024).


\bibitem{Poole2024a}
C. Poole, T. M. Graham, M. A. Perlin, M. Otten, and M. Saffman, 
Architecture for fast implementation of qLDPC codes with optimized Rydberg gates, 
arXiv:2404.18809.










\bibitem{Pino2021a}
J. M. Pino, J. M. Dreiling, C. Figgatt, J. P. Gaebler, S. A. Moses, M. S. Allman, C. H. Baldwin, 
M. Foss-Feig, D. Hayes, K. Mayer, C. Ryan-Anderson, and B. Neyenhuis, 
Demonstration of the trapped-ion quantum CCD computer architecture, 
Nature \textbf{592}, 209 (2021).

\bibitem{Ryan2021a}
C. Ryan-Anderson, J. G. Bohnet, K. Lee, D. Gresh, A. Hankin, J. P. Gaebler, D. Francois, A. Chernoguzov, D. Lucchetti, 
N. C. Brown, T. M. Gatterman, S. K. Halit, K. Gilmore, J. A. Gerber, B. Neyenhuis, D. Hayes, and R. P. Stutz, 
Realization of Real-Time Fault-Tolerant Quantum Error Correction, 
Phys. Rev. X \textbf{11}, 041058 (2021).

\bibitem{Postler2022a}
L. Postler, S. Heu{\ss}en, I. Pogorelov, M. Rispler, T. Feldker, M. Meth, C. D. Marciniak, R. Stricker, M. Ringbauer, 
R. Blatt, P. Schindler, M. M\"uller, and T. Monz, 
Demonstration of fault-tolerant universal quantum gate operations, 
Nature \textbf{605}, 675 (2022).

\bibitem{Moses2023a}
S. A. Moses, C. H. Baldwin, M. S. Allman, R. Ancona, L. Ascarrunz, C. Barnes \textit{et al}.,
A Race-Track Trapped-Ion Quantum Processor, 
Phys. Rev. X \textbf{13}, 041052 (2023).

\bibitem{Wang2024a}
Y. Wang, S. Simsek, T. M. Gatterman, J. A. Gerber, K. Gilmore, D. Gresh, N. Hewitt, C. V. Horst, M. Matheny, 
T. Mengle, B. Neyenhuis, and B. Criger, 
Fault-tolerant one-bit addition with the smallest interesting color code, 
Sci. Adv. \textbf{10}, eado9024 (2024).


\bibitem{Postler2024a}
L. Postler, F. Butt, I. Pogorelov, C. D. Marciniak, S. Heu{\ss}en, R. Blatt, 
P. Schindler, M. Rispler, M. M\"uller, and T. Monz, 
Demonstration of Fault-Tolerant Steane Quantum Error Correction, 
PRX Quantum \textbf{5}, 030326 (2024).





\bibitem{Self2024a}
C. N. Self, M. Benedetti, and D. Amaro, 
Protecting expressive circuits with a quantum error detection code, 
Nat. Phys. \textbf{20}, 219 (2024).

\bibitem{Yamamoto2024a}
K. Yamamoto, S. Duffield, Y. Kikuchi, and D. M. Ramo, 
Demonstrating Bayesian quantum phase estimation with quantum error detection, 
Phys. Rev. Res. \textbf{6}, 013221 (2024).

\bibitem{Silva2024a}
M. P. da Silva, C. Ryan-Anderson, J. M. Bello-Rivas,1 A. Chernoguzov, J. M. Dreiling, 
C. Foltz, F. Frachon, 
J. P. Gaebler, T. M. Gatterman, L. Grans-Samuelsson, D. Hayes, N. Hewitt, 
J. Johansen, D. Lucchetti, 
M. Mills, S. A. Moses, B. Neyenhuis, A. Paz, J. Pino, P. Siegfried, J. Strabley, 
A. Sundaram, 
D. Tom, S. J. Wernli, M. Zanner, R. P. Stutz, and K. M. Svore, 
Demonstration of logical qubits and repeated error correction with better-than-physical error rates, 
arxiv:2404.02280 (2024).

\bibitem{Ryan2024a}
C. Ryan-Anderson, N. C. Brown, C. H. Baldwin, J. M. Dreiling, C. Foltz, J. P. Gaebler, T. M. Gatterman, 
N. Hewitt, C. Holliman, C. V. Horst, J. Johansen, D. Lucchetti, T. Mengle, M. Matheny, Y. Matsuoka, 
K. Mayer, M. Mills, S. A. Moses, B. Neyenhuis, J. Pino, P. Siegfried, R. P. Stutz, J. Walker, and D. Hayes, 
High-fidelity teleportation of a logical qubit using transversal gates and lattice surgery,
Science \textbf{385}, 1327 (2024).






\bibitem{Dasu2025a}
S. Dasu, S. Burton, K. Mayer, D. Amaro, J. A. Gerber, K. Gilmore,
D. Gresh, D. DelVento, A. C. Potter, and D. Hayes, 
Breaking even with magic: demonstration of a high-fidelity logical non-Clifford gate, 
arXiv:2506.14688.


\bibitem{Daguerre2025a}
L. Daguerre, R. Blume-Kohout, N. C. Brown, D. Hayes, and I. H. Kim, 
Experimental demonstration of high-fidelity logical magic states from code switching, 
arXiv:2506.14169.


\bibitem{Hughes2025a}
A. C. Hughes, R. Srinivas, C. M. L\"oschnauer, H. M. Knaack, 
R. Matt, C. J. Ballance, M. Malinowski, T. P. Harty, and R. T. Sutherland,
Trapped-ion two-qubit gates with $>$ 99.99\% fidelity without ground-state cooling,
arXiv:2510.17286.

\bibitem{Ransford2025a}
A. Ransford, M. S. Allman, J. Arkinstall, J.P. Campora III, S. F. Cooper, R. D. Delaney \textit{et al}., 
Helios: A 98-qubit trapped-ion quantum computer,
arXiv:2511.05465.





\bibitem{Graham2022a}
T. M. Graham, Y. Song, J. Scott, C. Poole, L. Phuttitarn, K. Jooya, P. Eichler, X. Jiang, A. Marra, 
B. Grinkemeyer, M. Kwon, M. Ebert, J. Cherek, M. T. Lichtman, M. Gillette, J. Gilbert, 
D. Bowman, T. Ballance, C. Campbell, E. D. Dahl, O. Crawford, N. S. Blunt, B. Rogers, 
T. Noel, and M. Saffman, 
Multi-qubit entanglement and algorithms on a neutral-atom quantum computer, 
Nature \textbf{604}, 457 (2022).


\bibitem{Bluvstein2022a}
D. Bluvstein, H. Levine, G. Semeghini, T. T. Wang, S. Ebadi, M. Kalinowski, A. Keesling, N. Maskara, 
H. Pichler, M. Greiner, V. Vuletić, and M. D. Lukin, 
A quantum processor based on coherent transport of entangled atom arrays, 
Nature \textbf{604}, 451 (2022).


\bibitem{Evered2023a}
S. J. Evered, D. Bluvstein, M. Kalinowski, S. Ebadi, T. Manovitz, H. Zhou, S. H. Li, A. A. Geim, T. T. Wang, 
N. Maskara, H. Levine, G. Semeghini, M. Greiner, V. Vuletić, and M. D. Lukin, 
High-fidelity parallel entangling gates on a neutral-atom quantum computer, 
Nature \textbf{622}, 268 (2023).



\bibitem{Bluvstein2024a}
D. Bluvstein, S. J. Evered, A. A. Geim, S. H. Li, H. Zhou, T. Manovitz, S. Ebadi, M. Cain, 
M. Kalinowski, D. Hangleiter, J. P. B. Ataides, N. Maskara, I. Cong, X. Gao, P. S. Rodriguez, 
T. Karolyshyn, G. Semeghini, M. J. Gullans, M. Greiner, V. Vuleti\'{c}, and M. D. Lukin, 
Logical quantum processor based on reconfigurable atom arrays, 
Nature \textbf{626}, 58 (2024).


\bibitem{Muniz2025a}
J. A. Muniz, M. Stone, D. T. Stack, M. Jaffe, J. M. Kindem, L. Wadleigh \textit{et al}., 
High-Fidelity Universal Gates in the 171Yb Ground-State Nuclear-Spin Qubit,
PRX Quantum \textbf{6}, 020334 (2025).


\bibitem{Bluvstein2025a}
D. Bluvstein, A. A. Geim, S. H. Li, S. J. Evered, J. P. B. Ataides, G. Baranes \textit{et al}., 
Architectural mechanisms of a universal fault-tolerant quantum computer,
arXiv:2506.20661.


\bibitem{Rodriguez2025a}
P. S. Rodriguez, J. M. Robinson, P. N. Jepsen, Z. He, C. Duckering, C. Zhao \textit{et al}., 
Experimental demonstration of logical magic state distillation,
Nature \textbf{645}, 620 (2025).

\bibitem{Chiu2025a}
N.-C. Chiu, E. C. Trapp, J. Guo, M. H. Abobeih, L. M. Stewart, S. Hollerith \textit{et al}., 
Continuous operation of a coherent 3,000-qubit system, 
Nature \textbf{646}, 1075 (2025).

\bibitem{Manetsch2025a}
H. J. Manetsch, G. Nomura, E. Bataille, X. Lv, K. H. Leung, and M. Endres,
A tweezer array with 6,100 highly coherent atomic qubits, 
Nature \textbf{647}, 60 (2025).





\bibitem{Infleqtion2025a}
Infleqtion and Collaborators, 
Demonstration of a Logical Architecture Uniting Motion and In-Place Entanglement:
Shor's Algorithm, Constant-Depth CNOT Ladder, and Many-Hypercube Code, 
arXiv:2509.13247.




\bibitem{Bravyi2010a}
S. Bravyi, D. Poulin, and B. Terhal, 
Tradeoffs for Reliable Quantum Information Storage in 2D Systems, 
Phys. Rev. Lett. \textbf{104}, 050503 (2010).

\bibitem{Baspin2022a}
N. Baspin and A. Krishna, 
Quantifying Nonlocality: How Outperforming Local Quantum Codes Is Expensive, 
Phys. Rev. Lett. \textbf{129}, 050505 (2022).









\bibitem{Krishna2021a}
A. Krishna and D. Poulin, 
Fault-Tolerant Gates on Hypergraph Product Codes, 
Phys. Rev. X \textbf{11}, 011023 (2021).

\bibitem{Cohen2022a}
L. Z. Cohen, I. H. Kim, S. D. Bartlett, and B. J. Brown, 
Low-overhead fault-tolerant quantum computing using long-range connectivity, 
Sci. Adv. \textbf{8}, eabn1717 (2022).

\bibitem{Quintavalle2023a}
A. O. Quintavalle, P. Webster, and M. Vasmer, 
Partitioning qubits in hypergraph product codes to implement logical gates, 
Quantum \textbf{7}, 1153 (2023).



\bibitem{Williamson2024a}
D. J. Williamson and T. J. Yoder,
Low-overhead fault-tolerant quantum computation by gauging logical operators, 
arXiv:2410.02213.


\bibitem{Swaroop2024a}
E. Swaroop, T. Jochym-O'Connor, and T. J. Yoder, 
Universal adapters between quantum LDPC codes, 
arXiv:2410.03628.


\bibitem{Cross2024a}
A. Cross, Z. He, P. Rall, and T. Yoder, 
Improved QLDPC Surgery: Logical Measurements and Bridging Codes, 
arXiv:2407.18393.




\bibitem{Patra2024a}
A. Patra and A. Barg, 
Targeted Clifford Logical Gates for Hypergraph Product Codes, 
arXiv:2411.17050.

\bibitem{Malcolm2025a}
A. J. Malcolm, A. N. Glaudell, P. Fuentes, D. Chandra, A. Schotte, C. DeLisle, R. Haenel, 
A. Ebrahimi, J. Roffe, A. O. Quintavalle, S. J. Beale, N. R. Lee-Hone, and S. Simmons, 
Computing Efficiently in QLDPC Codes, 
arXiv:2502.07150.




\bibitem{Zhang2025a}
G. Zhang and Y. Li,
Time-Efficient Logical Operations on Quantum Low-Density Parity Check Codes, 
Phys. Rev. Lett. 134, 070602 (2025).




\bibitem{Xu2025a}
Q. Xu, H. Zhou, G. Zheng, D. Bluvstein, 
J. P. B. Ataides, M. D. Lukin, and L. Jiang, 
Fast and Parallelizable Logical Computation with Homological Product Codes, 
Phys. Rev. X \textbf{15}, 021065 (2025).




\bibitem{Ide2025a}
B. Ide, M. G. Gowda, P. J. Nadkarni, and G. Dauphinais, 
Fault-Tolerant Logical Measurements via Homological Measurement, 
Phys. Rev. X \textbf{15}, 021088 (2025).





\bibitem{Cowtan2025a}
A. Cowtan, Z. He, D. J. Williamson, and T. J. Yoder,
Parallel Logical Measurements via Quantum Code Surgery, 
arXiv:2503.05003.




\bibitem{He2025a}
Z. He, A. Cowtan, D. J. Williamson, and T. J. Yoder,
Extractors: QLDPC Architectures for Efficient Pauli-Based Computation, 
arXiv:2503.10390.






\bibitem{Yoder2025a}
T. J. Yoder, E. Schoute, P. Rall, E. Pritchett, J. M. Gambetta,
A. W. Cross, M. Carroll, and M. E. Beverland,
Tour de gross: A modular quantum computer based on bivariate
bicycle codes,
arXiv:2506.03094.


\bibitem{Baspin2025a}
N. Baspin, L. Berent, and L. Z. Cohen,
Fast surgery for quantum LDPC codes, 
arXiv:2510.04521.


\bibitem{Xu2025b}
Q. Xu, H. Zhou, D. Bluvstein, M. Cain, M.
Kalinowski, J. Preskill, M. D. Lukin, and N. Maskara, 
Batched high-rate logical operations for quantum LDPC codes, 
nonarXiv:
2510.06159.





\bibitem{Zheng2025a}
G. Zheng, L. Jiang, and Q. Xu, 
High-Rate Surgery: towards constant-overhead logical operations, 
arXiv:2510.08523.

\bibitem{Cowtan2025b}
A. Cowtan, Z. He, D. J. Williamson, and T. J. Yoder, 
Fast and fault-tolerant logical measurements:
Auxiliary hypergraphs and transversal surgery,
arXiv:2510.14895.



\bibitem{Zhang2025b}
G. Zhang, Y. Zhu, and Y. Li,
Accelerating Fault-Tolerant Quantum Computation with Good qLDPC Codes,
arXiv:2510.19442.





\bibitem{Pecorari2025a}
L. Pecorari, F. P. Guerci, H. Perrin, and G. Pupillo,
Addressable gate-based logical computation with quantum LDPC codes,
arXiv:2511.06124.

\bibitem{Webster2025a}
P. Webster, S. C. Smith, and L. Z. Cohen,
Explicit construction of low-overhead gadgets for gates on quantum LDPC codes, 
arXiv:2511.15989.


\bibitem{Tan2025a}
S. J. S. Tan, Y. Hong, T.-C. Lin, S, M. J. Gullans, and M.-H. Hsieh, 
Single-Shot Universality in Quantum LDPC Codes via Code-Switching,
arXiv:2510.08552.




\bibitem{Tamiya2025a}
S. Tamiya, M. Koashi, and H. Yamasaki, 
Fault-tolerant quantum computation with polylogarithmic time and constant space overheads, 
Nat. Phys. (2025). https://doi.org/10.1038/s41567-025-03102-5.









\bibitem{Yamasaki2024a}
H. Yamasaki and M. Koashi, 
Time-Efficient Constant-Space-Overhead Fault-Tolerant Quantum Computation, 
Nat. Phys. \textbf{20}, 247 (2024).


\bibitem{Goto2024a}
H. Goto, 
High-performance fault-tolerant quantum computing with many-hypercube codes, 
Sci. Adv. \textbf{10}, eadp6388 (2024).


\bibitem{Yoshida2025a}
S. Yoshida, S. Tamiya, and H. Yamasaki, 
Concatenate codes, save qubits, 
npj Quant. Inf. \textbf{11}, 88 (2025).


\bibitem{Litinski2025a}
D. Litinski, 
Blocklet concatenation: 
Low-overhead fault-tolerant protocols for fusion-based quantum computation, 
arXiv:2506.13619.

\bibitem{Liu2025a}
P. Liu, M. Xu, H. Zhou, H. Wang, U. A. Acar, and Y. Shi, 
ConiQ: Enabling Concatenated Quantum Error Correction on Neutral Atom Arrays,
arXiv:2508.05779.

\bibitem{Kanomata2025a}
K. Kanomata and H. Goto, 
Fault-tolerant quantum computing with a high-rate symplectic double code, 
arXiv:2509.15457.


\bibitem{Nakai2025a}
R. Nakai and H. Goto, 
Subsystem many-hypercube codes: High-rate concatenated codes with low-weight syndrome measurements,
arXiv:2510.04526.


\bibitem{Berthusen2025a}
N. Berthusen and E. Durso-Sabina,
Simple logical quantum computation with concatenated symplectic double codes, 
arXiv:2510.18753.



\bibitem{comment_decoder}
The decoder proposed in Ref.~\citenum{Goto2024a} is dedicated to 
$[[6^L,4^L,2^L]]$ MHC codes ($L$ is the concatenation level), 
but this can be extended to any combination of $[[6,4,2]]$ and 
$[[4,2,2]]$ straightforwardly.



\bibitem{Chao2018a}
R. Chao and B. W. Reichardt, Quantum Error Correction with Only Two Extra Qubits, Phys. Rev. Lett. \textbf{121}, 050502 (2018).


\bibitem{Reichardt2018a}
B. W. Reichardt, 
Fault-tolerant quantum error correction 
for Steane's seven-qubit color code with few or no extra qubits,
arXiv:1804.06995 (2018).




\bibitem{Steane1997a}
A. M. Steane, 
Active Stabilization, Quantum Computation, and Quantum State Synthesis, 
Phys. Rev. Lett. \textbf{78}, 2252 (1997).





\bibitem{Gidney2021a}
C. Gidney, 
Stim: a fast stabilizer circuit simulator, 
Quantum \textbf{5}, 497 (2021).



\bibitem{Knill2005b}
E. Knill, 
Scalable quantum computing in the presence of large detected-error rates, 
Phys. Rev. A \textbf{71}, 042322 (2005).



\end{thebibliography}
\end{document}